\documentclass[aps,prd,showpacs,twocolumn]{revtex4}
\usepackage[parfill]{parskip}    % Activate to begin paragraphs with an empty line rather than an indent
\usepackage{graphicx}
\usepackage{amssymb}
\usepackage{epstopdf}
\usepackage{color}
\usepackage{float}
\usepackage{ulem}
\usepackage{graphicx}
\usepackage{pdfpages}
\usepackage{amsmath}
\usepackage{physics}

\usepackage[colorlinks=true,citecolor=blue,urlcolor=magenta,breaklinks]{hyperref}
\usepackage{xcolor}     % for colors

\begin{document}

\title{A topological invariant in the context of the loop representation of the massive Kalb-Ramond-Klein-Gordon model}

\author{E. I\~niguez}
\affiliation{Departamento de F\'isica, Colegio de Ciencias e Ingenier\'ia, Universidad San Francisco de Quito (USFQ),  Quito , Ecuador\\}

\author{M. Freire}
\affiliation{Departamento de F\'isica, Colegio de Ciencias e Ingenier\'ia, Universidad San Francisco de Quito (USFQ),  Quito , Ecuador\\}

\author{L. Leal }
\affiliation{Centro de F\'isica Te\'orica y Computacional,\\ Escuela de F\'isica, Facultad de Ciencias, Universidad Central de Venezuela, Caracas 1050, Venezuela\\}

\author{E. Contreras }
\email{ernesto.contreras@ua.es}
\affiliation{Departamento de F\'{\i}sica Aplicada, Universidad de Alicante, Campus de San Vicente del Raspeig, E-03690 Alicante, Spain.\\}

%%%%%%%%%%%%%%%%%%%%%%%%%%%%%%%%%%
\begin{abstract}
We employ the Dirac procedure to quantize the self-dual massive Kalb-Ramond-Klein-Gordon model in $2+1$ dimensional spacetimes. The canonical fields are expressed in terms of $2$-surfaces and signed points, ensuring the automatic realization of the quantum algebra. As the duality rotation preserving the action can be implemented infinitesimally, we derive the conserved quantity that generates the transformation. Given that such a generator is a two dimensional topological quantity, its representation in terms of geometrical operators yields a two dimensional invariant (reminiscent of a projection of Gauss's law in electrodynamics), which encodes the same information of the well-known winding number. 
\end{abstract}
%%%%%%%%%%%%%%%%%%%%%%%%%%%%%%%%%%%%%%%%%%

\maketitle

%%%%%%%%%%%%%%%%%%%%%%%%%%%%%%%%%%%%%%%%%%%%%%%%%%%%%%%%%%%%%%%%%%%%%%%%
\section{Introduction}\label{intro}
Duality establishes profound connections between seemingly unrelated models or different aspects of the same model, often revealing unexpected relationships. One notable example is the duality in vacuum Maxwell equations, which are invariant under rotations between electric and magnetic fields \cite{DeserDualityTO}:

\[
\begin{pmatrix}
\vec{E}' \\
\vec{B}'
\end{pmatrix}
=
\begin{pmatrix}
\cos\theta & \sin\theta \\
-\sin\theta & \cos\theta
\end{pmatrix}
\begin{pmatrix}
\vec{E}\\
\vec{B}
\end{pmatrix}
\]
which follows from the infinitesimal transformations:
\begin{align*}
\delta \vec{E} &= \theta \nabla \times \vec{A} \\
\delta \vec{A} &= \theta \nabla^{-2} \nabla \times \vec{E}.
\end{align*}
This symmetry extends to theories involving massive and massless $p$-forms in $D$-dimensional spacetimes \cite{ContrerasPRD}. By Noether’s theorem, a conserved quantity must exist when the transformation is imposed at an infinitesimal level, particularly for self-dual models where the $p$-form sector maps into the $q$-form sector via the dual-Hodge operation. Such generators of duality transformations emerge as topological quantities, independent of the spacetime metric. When quantized in a geometric representation of $p$-surfaces, the operator associated with the duality generator becomes a link invariant. For massless $p$-form models, this corresponds to a generalization of the Gauss linking number, whereas for massive $p$-form fields, it is interpreted as the intersection number between open $p$-surfaces \cite{ContrerasPRD}.

In this work, we explore the duality generator in the massive Kalb-Ramond-Klein-Gordon (KRKG) theory in $2+1$-dimensional spacetime \cite{almeida2004dualization, Deser1997NForms, Medeiros_2001, Casini2002, Arias:2002cd, Camacaro:1996fa}. The reason for considering models in $2+1$ dimensions is twofold. First, unlike previous studies that analyzed general massive and massless $p$-form models, our work focuses explicitly on the self-dual massive KRKG model in $2+1$ dimensions, ensuring a more concrete and physically motivated analysis. A key novelty of our approach is the introduction of a two-dimensional topological invariant formulated in terms of 2-surfaces and signed points. This invariant is directly related to Gauss’s law in electrodynamics, providing a novel interpretation absent in prior discussions of duality in massive $p$-form theories. Moreover, we establish a rigorous connection between our invariant and the winding number, which was not explicitly explored in previous studies. This offers a more intuitive geometric and physical understanding of the topological structures underlying the model. Secondly, in recent years, $2+1$ dimensional quantum field theories have attracted considerable interest due to their relevance in condensed matter systems such as topological insulators, graphene, and quantum Hall fluids. These systems naturally host effective theories in lower dimensions, where topology and duality play an enhanced role. In particular, exact results and geometric interpretations become more accessible, as seen in studies of Levinson's theorem for the Dirac and Klein-Gordon equations in two dimensions~\cite{Dong1998,Dong1999}. These works reveal how lower-dimensional field theories admit subtle spectral and topological phenomena, including half-bound states and nontrivial scattering behavior. Furthermore, the adaptation of angular momentum algebra to low dimensions, as discussed in~\cite{Gu2002}, illustrates how group-theoretical structures persist and support exact solutions. While our model is not directly tied to condensed matter applications, the study of a topological invariant in the massive KRKG theory in $2+1$ dimensions offers a geometrically inspired perspective that may contribute to future connections with low-dimensional quantum systems.

This work is structured as follows. In the next section, we introduce the KRKG model in $2+1$ dimensions, constructing its Hamiltonian and duality generator. In Section \ref{representation}, we review the main aspects of the geometric representation used in this work. Section \ref{geometrical} details the quantization of the model using 2-surfaces and signed-point operators, followed by an interpretation of the results. The final section presents concluding remarks and potential future directions.

\section{Massive Kalb-Ramond-Klein-Gordon model in $2+1$ dimensions}\label{krkg}
The action of massive KRKG theory in $2+1$ dimensional spacetimes, reads
\begin{eqnarray}\label{action}
    S &=& \int d^{3}x \bigg(\frac{1}{12}F_{\mu\nu\lambda}F^{\mu\nu\lambda} - \frac{m^{2}}{4}A_{\mu\nu}A^{\mu\nu}\\
    &+&\frac{1}{2}\partial_{\mu}\phi\partial^{\mu}\phi - \frac{m^{2}}{2}\phi^{2}\bigg)
\end{eqnarray}
where 
\begin{equation}
    F_{\mu\nu\lambda} = \frac{1}{2}\partial_{[\mu}A_{\nu\lambda]} = \partial_{\mu}A_{\nu\lambda} + \partial_{\nu}A_{\lambda\mu} + \partial_{\lambda}A_{\mu\nu}
\end{equation}
and $A_{\mu\nu}$ is a totally antisymmetric tensor, and $\phi$ is a scalar field. Note that, the action (\ref{action}) corresponds to a self--dual massive model in the Hodge operation sense. In this case, as the Kalb-Ramond field is a $2$-form and the scalar field is a $0$-form, the relation $D-1=p+q$ holds, as expected.

Making variations on the action, we arrive at
\begin{align}
    \partial_{\lambda}F^{\mu\nu\lambda} + m^{2}A^{\mu\nu} &= 0 \label{eq:KalbRamond}\\
    \partial_{\mu}\partial^{\mu}\phi + m^{2}\phi &= 0 .\label{eq:KleinGordon}
\end{align}
Equations \eqref{eq:KalbRamond} and \eqref{eq:KleinGordon} can be expressed as
\begin{align}
    \partial_{i}E^{ij} + m^{2}A^{j0} &= 0\label{e10}\\
    \dot{E}^{ij} + m^{2}A^{ij} &=0\label{e2}\\
    E^{ij} - \partial_{i}A^{j0} - \partial_{j}A^{0i} - \dot{A}^{ij} &=0\label{e3}\\
    D - \dot{\phi} &=0\label{e4}\\
    \partial_{i}\phi + \varepsilon^{ij}H^{j} &=0\label{e5}\\
    \dot{D} + \varepsilon^{ij}\partial_{i}H^{j} + m^{2}\phi &=0\label{e6}.
\end{align}
where the dot stand for the partial derivative $\partial_{0}$
and 
\begin{align}
    E^{ij} &= F^{0ij}\\
    \dot{\phi} &= D\\
    \partial_{i}\phi &= -\varepsilon^{ij}H^{j},
\end{align}

Note that, if we define the quantities
\begin{align}
    \tilde{F} &= \begin{pmatrix} E^{ij} \\ D \\ H^{i} \end{pmatrix}
\end{align}
and
\begin{align}
    \tilde{A} &= \begin{pmatrix} m\varepsilon^{ij}\phi \\ \frac{m}{2}\varepsilon^{ij}A^{ij} \\ -mA^{i0} \end{pmatrix},
\end{align}
is clear that the set of Eqs. (\ref{e10})-(\ref{e6}) are invariant under duality rotations, namely
\[\begin{pmatrix} \tilde{F}' \\ \tilde{A}' \end{pmatrix} = \begin{pmatrix}
\cos\theta & \sin\theta \\
-\sin\theta & \cos\theta
\end{pmatrix} \begin{pmatrix} \tilde{F} \\ \tilde{A} \end{pmatrix}\]

It should be highlighted that this symmetry of the field equations is also a symmetry of the action when applied to the dynamic fields of the theory. Additionally, since it can be implemented infinitesimally, there exists a conserved quantity by Noether's theorem. In this work, we are interested in quantizing the theory, so the next step in the program is to construct the Hamiltonian and the Noether charge that generates the transformations leaving the first-order action invariant. In order to achieve this, we follow the Dirac procedure \cite{ContrerasPRD}. As a first step, we explicitly separate space and time to identify the velocities, namely
\begin{eqnarray}
S&=&\int d^{3}x\bigg(\frac{1}{4}F_{ij0}F^{ij0}-\frac{m^{2}}{2}A_{i0}A^{i0}\nonumber
-\frac{m^{2}}{4}A_{ij}A^{ij}\\
&&+\frac{1}{2}\dot{\phi}^{2}+\frac{1}{2}\partial_{i}\phi\partial^{i}\phi-\frac{m^{2}}{2}\phi^{2}\bigg)    
\end{eqnarray}
From the above expression, we obtain the conjugate momentum
\begin{eqnarray}
E^{ij}&=&\frac{\partial\mathcal{L}}{\partial \dot{A}_{ij}}=\dot{A}_{ij}+\partial_{i}A_{j0}+\partial_{j}A_{0i}\\
D&=&\frac{\partial \mathcal{L}}{\partial\dot{\phi}}=\dot{\phi}
\end{eqnarray}
Now, the Hamiltonian 
\begin{eqnarray}
\mathcal{H}=\int d^{2}x\left(\frac{1}{2}E^{ij}\dot{A}_{ij}+D\dot{\phi}-\mathcal{L}\right)    
\end{eqnarray}
can be written as

\begin{eqnarray}
    \label{eq:Hamiltonian}
        \mathcal{H} &=& \int d^{2}x\bigg(\frac{1}{4}E^{ij}E^{ij} + \frac{1}{2m^{2}}\partial_{i}E^{ik}\partial_{j}E^{jk} +\nonumber\\
       && \frac{m^{2}}{4}A_{ij}A^{ij} + \frac{1}{2}D^{2} + \frac{1}{2}H^{i}H^{i} + \frac{m^{2}}{2}\phi^{2}\bigg).
\end{eqnarray}
It is worth mentioning that during the process arise two second class constraints, namely
\begin{align}
    &\Pi^{0 i} = 0,\label{eq:const1}\\
    &\partial_{i}E^{ij} + m^{2}A^{j0}=0, \label{eq:const2}
\end{align}
so the canonical algebra of the theory is given by the Dirac brackets \cite{Dirac1964}
\begin{equation}
    \label{eq:DiracBrackets}
    \begin{split}
        \{A_{ij}(\vec{x}),E^{mn}(\vec{y})\}^{*} &= \delta_{ij}^{mn}\delta^{2}(\vec{x}-\vec{y})\\
  \{\phi(\vec{x}),D(\vec{y})\}^{*} &= \delta^{2}(\vec{x}-\vec{y})\\
        \{A_{ij}(\vec{x}),A_{mn}(\vec{y})\}^{*} &= \{E^{ij}(\vec{x}),E^{mn}(\vec{y})\}^{*} = 0\\
        \{\phi(\vec{x}),\phi(\vec{y})\}^{*} &= \{D(\vec{x}),D(\vec{y})\}^{*} = 0.
    \end{split}
\end{equation}
The first order action reads\\
\begin{eqnarray}
    S &=& \int d^{3}x \bigg(\frac{1}{2}E^{ij}\dot{A_{ij}}+D\dot{\phi}-\frac{1}{4}E^{ij}E^{ij}-\frac{m^{2}}{4}A_{ij}A^{ij}\nonumber\\
   &-&\frac{1}{2m^{2}}\partial_{i}E^{ik}\partial_{j}E^{jk}
  - \frac{1}{2}D^{2}-\frac{1}{2}H^{i}H^{i} - \frac{1}{2}m^{2}\phi^{2}
  \bigg)
\end{eqnarray}
which, up to a total time derivative term is invariant under the following duality transformation
\begin{align}
    \delta E^{ij} &= \theta m\varepsilon^{ij}\phi\label{dt1}\\
    \delta D &= \theta \frac{m}{2}\varepsilon_{ij}A^{ij}\label{dt2}\\
    \delta H^{i}&=\frac{\theta}{m}\partial_{j}E^{ji}\\
    \delta \phi &= -\frac{\theta}{2m}\varepsilon_{ij}E^{ij}\label{dt3}\\
    \delta A^{ij} &= -\frac{\theta}{m}\varepsilon^{ij}D\label{dt4}\\
    \delta(\partial_{j}E^{ji})&=-\theta mH^{i},
\end{align}
Indeed, making variations on the action we arrive at
\begin{eqnarray}
\delta_{\theta}S=\theta\int d^{3}x\frac{d}{dt}\bigg(\frac{m}{2}\phi\varepsilon_{ij}A^{ij}-\frac{1}{2m}\varepsilon_{ij}E^{ij}D\bigg)    
\end{eqnarray}
Then, under the equations of motion the first order action change
\begin{eqnarray}
\delta S&=&\int d^{3}x\frac{d}{dt}\bigg(\frac{1}{2}E^{ij}\delta A_{ij}+D\delta\phi\bigg)\\
&=&-\theta\int d^{3}x\frac{d}{dt}\bigg(\frac{1}{m}\varepsilon_{ij}E^{ij}D\bigg)    
\end{eqnarray}
After equating both terms the first order the Noether charge that generates these transformations is
\begin{equation}\label{GD}
    G = \theta\int d^{2}x \left(\frac{m}{2}\varepsilon_{ij}\phi A^{ij} + \frac{1}{2m}\varepsilon_{ij}DE^{ij}\right),
\end{equation}
as can be checked by computing the Poisson bracket between $G$ and the canonical variables. At this point, a couple of comments are in order. First, note that both terms correspond to topological invariants as they do not require the metric for their construction. The importance of this fact will be revealed next after introducing the $p$-surface and signed points representations. Second, the generator can be alternatively written as
\begin{equation}
    G = \theta\int d^{2}x \frac{m}{2}\varepsilon^{ij}A_{ij}\int_{x_{0}}^{x}dz^{k} \partial^{z}_{k}\phi + \theta\int d^{2}x\frac{1}{2m}\varepsilon_{ij}DE^{ij},
\end{equation}
where we have introduced the gradient of the scalar field explicitly for convenience (this point will be clarified later). 

\section{$p$-surface and signed points representations}\label{representation}
\begin{figure}[h!]
    \centering
    \includegraphics[width=0.5\columnwidth]{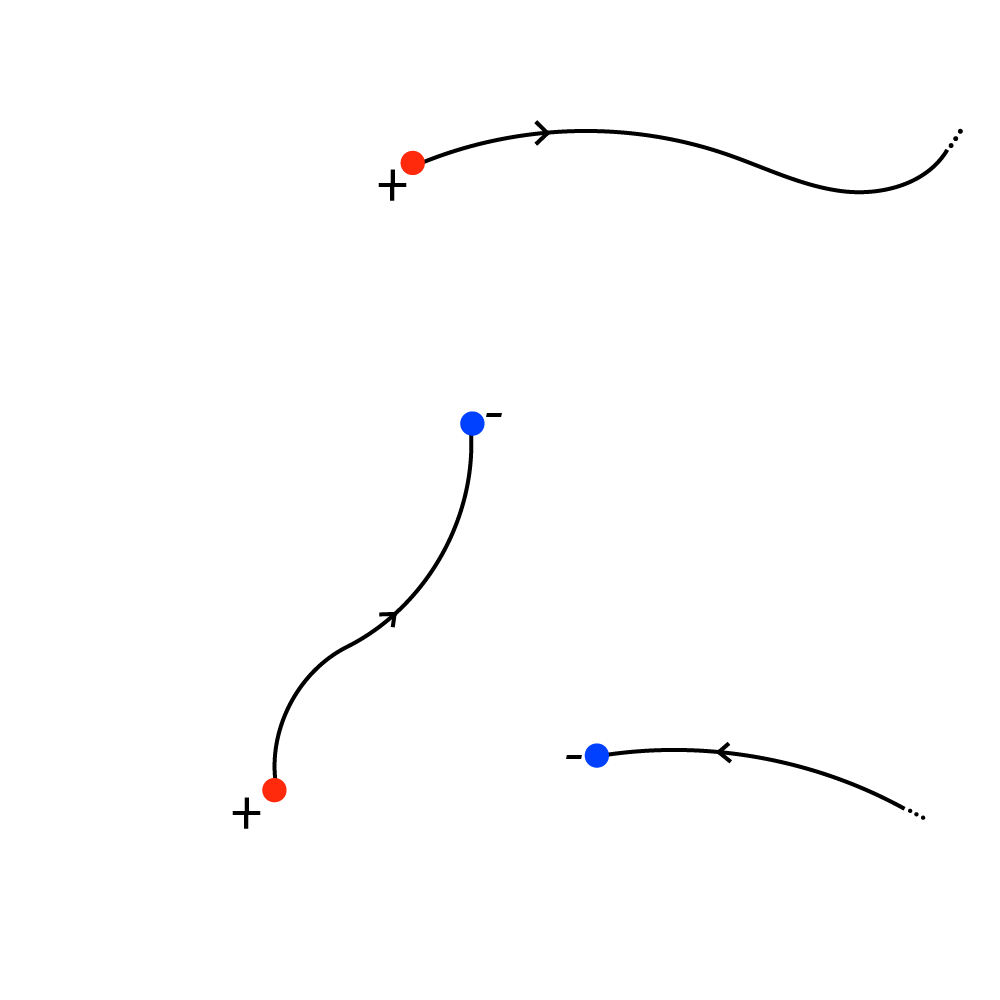}
    \caption{Points and anti-points as boundaries of Faraday's lines. Lines emerging from a point either converge to an anti-point or extend to infinity.}
    \label{fig:jodete}
\end{figure}

The Abelian $p$-surface space is nothing but a generalization of the Abelian path space and is defined as the set of a certain equivalence class of surfaces in a manifold, which we take as $R^{n}$ \cite{ContrerasPRD}. The equivalence classes of $p$-surfaces $\Sigma$ is the form-factor defined as
\begin{equation}
    T^{i_1 \ldots i_p}(\vec{x}, \Sigma) = \int_{\Sigma} d\Sigma_{\vec{y}}^{i_1\ldots i_p}   \delta^{D-1}(\vec{x} - \vec{y}),
\end{equation}
where $d\Sigma_{\vec{y}}^{i_1}\ldots _{i_p} $ is the surface element of $\Sigma$. Two surfaces $\Sigma$ and $\Sigma'$ will be considered equivalent if their form factor coincide. Now, we can define the open $p$-surface derivative $\delta_{i_{1}....i_{p}}(\vec{x})$ 
\begin{eqnarray}
\sigma^{i_{1}\cdots i_{p}}\delta_{i_{1}\cdots i_{p}}(\vec{x})\Psi[\Sigma]=\Psi[\Sigma\circ\sigma_{\vec{x}}]-\Psi[\Sigma],
\end{eqnarray}
that measures the change of the path-dependent functional $\Psi[\Sigma]$ when an infinitesimal surface of area $\sigma^{i_{1}\cdots i_{p}}$ is appended to its argument $\Sigma$ at the point $\vec{x}$. As an example of how this operator works, we calculate the open $p$-surface operator of the form factor. In this case, we have,
\begin{eqnarray}
T^{i_{1}...i_{p}}(\vec{x},\Sigma\circ\sigma_{\vec{y}})&=& \int\limits_{\Sigma\circ\sigma_{\vec{y}}}
d\Sigma_{\vec{y}}^{i_1\ldots i_p}   \delta^{D-1}(\vec{x} - \vec{y})\nonumber\\
&=&T^{i_{1}...i_{p}}(\vec{x},\Sigma)+\sigma^{i_{1}\ldots i_{p}}\delta^{D-1}(\vec{x}-\vec{y})\nonumber\\
\end{eqnarray}
from where
\begin{eqnarray}
\delta_{j_{1}\ldots j_{p}}(\vec{y})T^{i_{1}\ldots i_{p}}(\vec{x},\Sigma)=\delta^{i_{1}\ldots i_{p}}_{j_{1}\ldots j_{p}}\delta^{D-1}(\vec{x}-\vec{y}),
\end{eqnarray}
 where $\delta^{i_{1}\ldots i_{p}}_{j_{1}\ldots j_{p}}$ is the totally antisymmetric kroenecker delta. 
 As we will see later on, the index $p$ in the previous definitions will be associated with the index of the fields in our model. To be more precise, in our case, we are interested in the cases $p=2$ and $p=0$, which correspond to the dynamic differential forms in the massive KRKG model introduced in the previous section. However, what we have done so far does not include the case of $0$-surfaces, which is what we are going to develop next. In this case, we consider the set whose elements are unordered lists $X$ of ``signed'' points $x_s$, where $s = \pm$ \cite{LorenzoSignedPoints}. A list would be, for instance: $X = x^+_1, x^-_2, \ldots, x^+_s$. Note that, the points can be thought of as boundaries of oriented paths (maybe starting or ending at spatial infinity), inasmuch as loops can be seen as boundaries of 2-surfaces (see figure \ref{fig:jodete}). 

The form factor of a list is defined as
\begin{equation}
    T(x, X) = \sum_a s_a \delta^n(x - x_a), \label{eq:form-factor}
\end{equation}
where $s_a$ is the sign of the point at position $x_a$. Then we declare that two lists of points $X$ and $Y$ are equivalent if
\begin{equation}
    T(x, X) = T(x, Y). \label{eq:equivalence}
\end{equation}
Now we proceed as before. Considering functionals $\Psi[X]$ that depend on lists, we can define the operator $a(x_s)$ which acts on $\Psi[X]$ adding a signed point $x_{s}$ to the left of the argument $X$; that is
\begin{equation}
    a(x_{s})\Psi[X] = \Psi[x_{s}X].
\end{equation}
An immediate result from this definition is that for a list $Y$,
\begin{equation}
    a(Y)\Psi[X] = \Psi[Y^{-1}X]
\end{equation}
and successive applications of $a$ are taken to be
\begin{equation}
    a(X_{1})a(X_{2}) = a(X_{1}X_{2})
\end{equation}
for consistency. The inverse operator, $a^{-1}(x_{s})$ must act on $\Psi[X]$ by removing a signed point $x_{s}$. In the context of reduced lists, this is equivalent to adding a point of opposite sign $x_{-s}$, which makes $a^{-1}(x_{s})=a(x_{-s})$.

Let us consider an infinitesimal list of points $\delta Y$ defined as
\begin{equation}
    \delta Y = \{(x+u)_{s},x_{-s}\}
\end{equation}
up to first order in $u$. It is shown that we can also define an operator that measures the change of $\Psi[X]$ when an infinitesimal list of points is appended to its argument at the point $x$ by using
\begin{equation}
    \Psi[\delta YX] - \Psi[X] \equiv u^{i}\Delta_{i}(x)\Psi[X].
\end{equation}
Equivalently, this differential operator can be expressed in terms of the generators $a(x_{s})$ by the following identity (see \cite{LorenzoSignedPoints} for details)
\begin{equation}
    \Delta_i(x) \equiv a(x) \frac{\partial}{\partial x_i} a(x)^{-1},
\end{equation}
\\

Following the above prescription it can be shown that the signed point derivative of the form factor in Eq. (\ref{eq:form-factor}) leads to
\begin{align}
    \Delta_{i}(\vec{x})T(\vec{y},X) 
    =\pdv{}{x^{i}}\delta^{2}(\vec{x}-\vec{y}).
\end{align}

In the next section, we apply the canonical quantization rules to the massive KRKG system introduced in the previous section. In addition, we will use the operators defined here as a representation of the canonical quantum algebra. \\
\section{Generator of duality and geometrical representation}\label{geometrical}
In this section, we realize the generator of duality (\ref{GD}) in terms of $2$-surface space and signed points. To this end, let us start with the canonical quantization procedure, namely \cite{Dirac1964}
\begin{enumerate}
    \item The canonical variables become operators that act on vectors on a Hilbert Space:
    \begin{equation*}
        \begin{split}
            E^{ij}&\longrightarrow\hat{E}^{ij}\\
            A_{ij}&\longrightarrow\hat{A}_{ij}\\
            \phi&\longrightarrow\hat{\phi}\\
            D&\longrightarrow\hat{D}
        \end{split}
    \end{equation*}
    The constraints in \eqref{eq:const1} and \eqref{eq:const2} define the physical sector of the Hilbert Space.
    \item The commutator algebra is inherited from the Dirac brackets in \eqref{eq:DiracBrackets}, namely
  \begin{align}
[\hat{A}_{ij}(\vec{x}),\hat{E}^{mn}(\vec{y})] &=i \delta_{ij}^{mn}\delta^{2}(\vec{x}-\vec{y})\\
[\hat{\phi}(\vec{x}),\hat{D}(\vec{y})] &= i\delta^{2}(\vec{x}-\vec{y})\\
[\hat{A}_{ij}(\vec{x}),\hat{A}_{mn}(\vec{y})] &= [\hat{E}^{ij}(\vec{x}),\hat{E}^{mn}(\vec{y})] = 0\\
[\hat{\phi}(\vec{x}),\hat{\phi}(\vec{y})] &= [\hat{D}(\vec{x}),\hat{D}(\vec{y})] = 0.
  \end{align}
     
    \item The dynamics is given by the Schrödinger equation with the Hamiltonian in \eqref{eq:Hamiltonian}.
\end{enumerate}

Now, it can be shown the canonical algebra can be realized by the following representation of the canonical operators in terms of $2$-surface space and signed points operators as
\begin{align}
    \hat{E}^{ij}(\vec{x}) &= T^{ij}(\vec{x},\sigma)\, \label{e1}\\
    \Hat{A}_{ij}(\vec{x}) &= i\hat{\delta}_{ij}(\vec{x})\label{e2}\\
    \Hat{D}(\vec{x}) &= T(\vec{x},X)\, \label{e3}\\
    \partial_{i}\phi(\vec{x}) &= i\hat{\Delta}_{i}(\vec{x})\label{e4}.
\end{align}
In fact, let us consider a functional $\Psi[\Sigma,X]$ which depends on both 2-surfaces and list of signed points,
\begin{eqnarray}
[\hat{A}_{ij}(\vec{x}), \hat{E}^{kl}(\vec{y})]\,\Psi[\Sigma,X] 
&=& 
\bigl[i\,\delta_{ij}(\vec{x}),\,T^{kl}(\vec{y},\sigma)\bigr]\,\Psi[\Sigma,X] 
\nonumber \\
&=& 
i\,(\delta_{ij}(\vec{x})\,T^{kl}(\vec{y},\sigma))\,\Psi[\Sigma,X]
\nonumber \\
&=& 
i\,\delta_{ij}^{kl}\,\delta^{3}(\vec{x}-\vec{y})\,\Psi[\Sigma,X].
\end{eqnarray}
Similarly,
\begin{eqnarray}
[\partial_{i}\,\hat{\phi}(\vec{x}),\,\hat{D}(\vec{y})]\,\Psi[\Sigma,X] 
&=& 
i\,\bigl(\Delta_{i}(\vec{x})\,T(\vec{y},X)\bigr)\,\Psi[\Sigma,X]
\nonumber \\
&=& i
\frac{\partial}{\partial x^{i}}\delta^{2}(\vec{x}-\vec{y}).
\end{eqnarray}

\begin{figure}[h!]
    \centering
    \includegraphics[width=0.5\columnwidth]{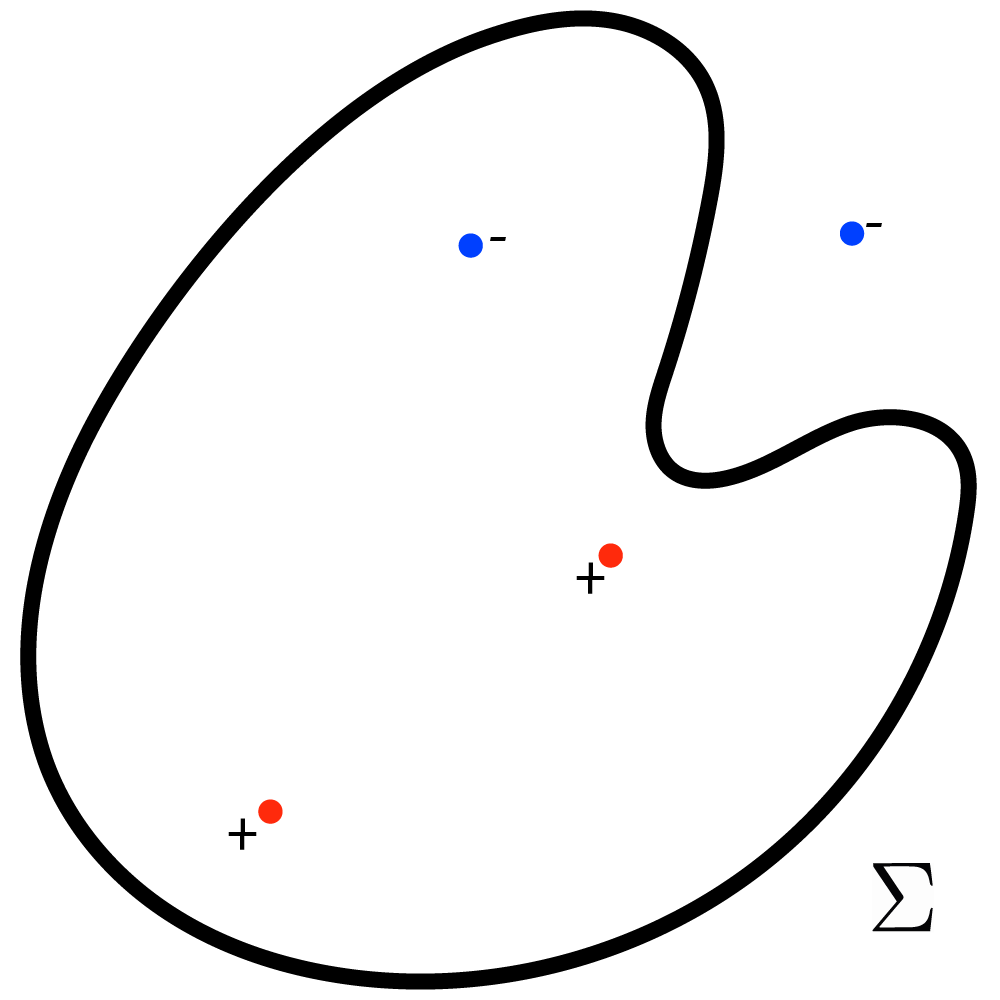}
    \caption{Points and antipoints within a \(2\)-surface \(\Sigma\)}
    \label{fig:point-antipoint-flux1}
\end{figure}

\begin{figure}[h!]
    \centering
    \includegraphics[width=0.5\columnwidth]{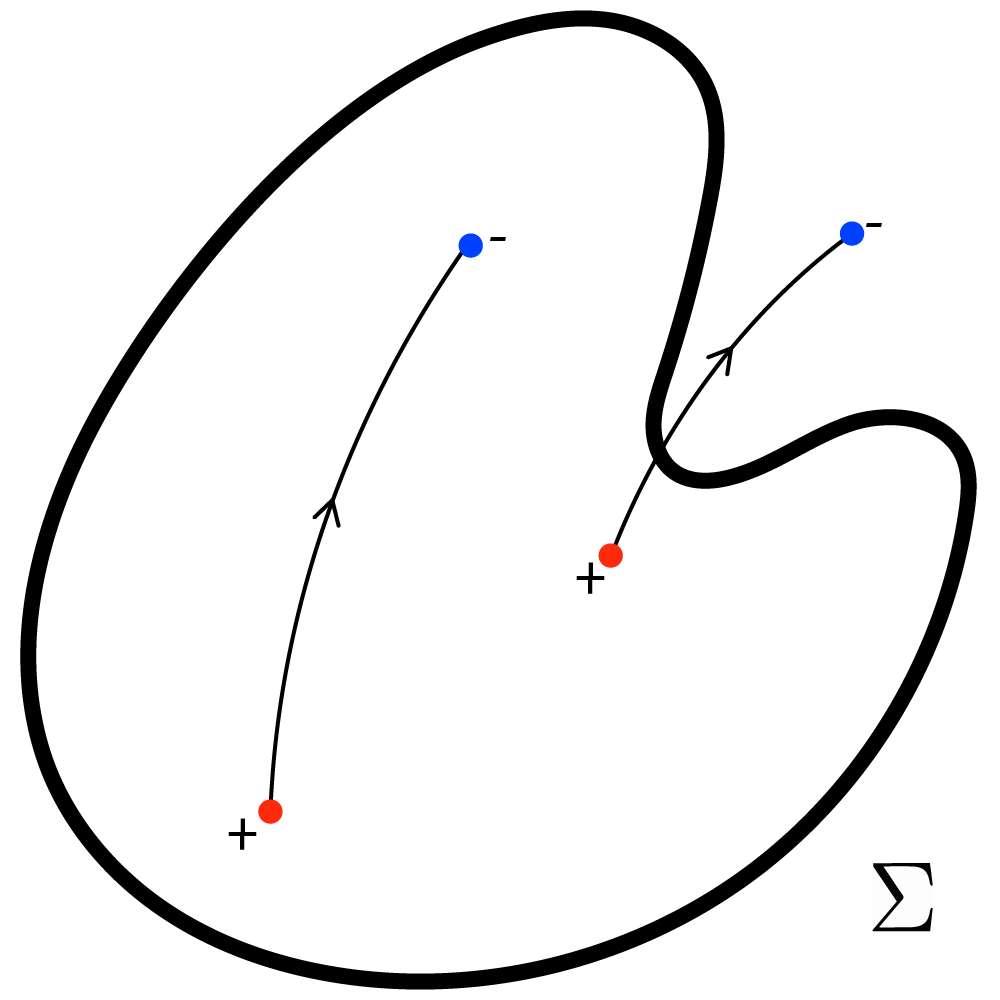}
    \caption{Points and antipoints with flux represented by attaching outgoing and incoming lines}
    \label{fig:point-antipoint-flux2}
\end{figure}

In terms of the realizations (\ref{e1})-(\ref{e4}), the generator reads
\begin{eqnarray}
    \Hat{G} &=& \int d^{2}x\; \frac{m}{2}\varepsilon^{ij}\hat{\delta}_{ij}(\vec{x})\int dz^{i}\hat{\Delta}_{i}(\vec{z})\nonumber\\
    &+& \int d^{2}x\; \varepsilon^{ij}\int d\Sigma_{\vec{y}}^{ij}\delta^{2}(\vec{x}-\vec{y})\sum_{a}s_{a}\delta^{2}(\vec{x}-\vec{z_{a}})\label{eq:GeneratorQuantum}
\end{eqnarray}
As we previously discussed, the duality generator is a topological quantity since it does not depend on the metric for its construction. Now, as its quantum counterpart is written in terms of surfaces and signed points, it must account for the linking between them. This fact is manifested in the second term which vanishes when a signed point lies outside $\Sigma$. To be more precise,  
\begin{eqnarray}\label{2dGL}
&&\int d^{2}x\; \varepsilon^{ij}\int d\Sigma_{\vec{y}}^{ij}\delta^{2}(\vec{x}-\vec{y})\sum_{a}s_{a}\delta^{2}(\vec{x}-\vec{z_{a}})=\\
&&=\sum_{a}s_{a}\int d\Sigma^{ij}_{\vec{y}}\delta^{2}(\vec{y}-\vec{z}_{a}).
\end{eqnarray}
Now,
\begin{eqnarray}
d\Sigma^{ij}=\frac{1}{2}dy^{i}\wedge dy^{j}.    
\end{eqnarray}
from where, 
\begin{eqnarray}
\varepsilon^{ij} d\Sigma^{ij}=dy^{2}    
\end{eqnarray}
With this in mind we obtain
\begin{eqnarray}
&&\sum\limits_{a}s_{a}\int \varepsilon^{ij}d\Sigma_{\vec{y}}\delta^{2}(\vec{y}-\vec{z}_{a})\\
 =&&\sum\limits_{a}s_{a}\int d^{2}y\delta^{2}(\vec{y}-\vec{z}_{a})
\end{eqnarray}
which is non-zero whenever the signed point located at $\vec{z}_{a}$ belong to the surface. We shall say that the sign of these points is $\tilde{s}_{a}$ so, finally the orignal expression van be wsritten as
\begin{eqnarray}
\int d^{2}x\; \varepsilon^{ij}\int d\Sigma_{\vec{y}}^{ij}\delta^{2}(\vec{x}-\vec{y})\sum_{a}s_{a}\delta^{2}(\vec{x}-\vec{z_{a}})=\sum_{a}\tilde{s}_{a}
\end{eqnarray}

where $\tilde{s}_{a}$ stands for the sign of the points on $\Sigma$. In figure \ref{fig:point-antipoint-flux1} we show a particular case where the number is $+1$.
Now, as discussed previously, we can interpret the signed points as the end points of oriented paths in a similar fashion to what occurs in electrodynamics. More precisely, as we can say that oriented paths are Faradays' lines that go from points to anti-points,  it can be seen that the second term of the generator of duality is a kind of flux as shown in figure \ref{fig:point-antipoint-flux2} so it represents a projected version of the Gauss law in electrodynamics.
This result highlights the intrinsic nature of the theory. In the free, massless case, all $p$-surfaces are closed, indicating the absence of monopoles (charges) within the framework. This is exemplified, for instance, in the Maxwell theory in $D=3+1$ dimensions, where Gauss's law, $\partial_{i}\hat{E}^{i}=0$, is automatically satisfied if the electric field is represented as the form factor of close paths. This follows from the fact that $\partial_{i}T^{i}(\vec{x},\gamma)=0$ holds whenever $\gamma$ is a loop. Conversely, in the free massive case (such as the Proca theory in $D=3+1$), an effective charge emerges, encoded in the temporal component of the vector potential: $\partial_{i}E^{i}=m^{2}A^{0}$. In this scenario, the path-based representation implies that the divergence of the form factor is nonzero, requiring open paths. In our context, the relevant geometric objects are open 2-surfaces and points/antipoints, which can be understood as open 0-surfaces. This result aligns with the findings of \cite{ContrerasPRD}, where it was demonstrated that the invariant corresponds to the intersection of an open $p$-surface and an open $q$-surface. The alternative interpretation presented here allows us to conclude that the invariant introduced in \cite{ContrerasPRD} is, in essence, a generalization of Gauss's law to $D$-dimensional spacetime.\\
Before concluding this section, we would like to discuss the connection between the invariant in (\ref{2dGL}) and the winding number. Let us consider a closed curve $\gamma$ enclosing a surface $\Sigma$. Suppose that within this surface there is a fluid with vortices, such that the enclosed vortices can be either clockwise or counterclockwise. The velocity field of a vortex at $\vec{a}$ is given by
\begin{eqnarray}
u^{j}=\kappa\varepsilon^{ij}\frac{(x^{i}-a^{i})}{|\vec{x}-\vec{a}|^{2}}
\end{eqnarray}
where $\kappa$ is the sign of the vortex ($+$ for clockwise and $-$ for counterclockwise, for example).\\
Now, suppose that $\gamma$ encloses $n$ vortices, which can be either clockwise or counterclockwise. The winding number around the points $\{a^{i}_{k}\}$ is given by
\begin{eqnarray}
\omega=\frac{1}{2\pi}\sum\limits_{k}\oint\limits_{\gamma}\kappa_{k}\varepsilon^{ij}\frac{(x^{i}-a^{i}_{k})}{|\vec{x}-\vec{a}_{k}|^{2}}dx^{j}
\end{eqnarray}
which, by applying Stokes' theorem, can be rewritten as
\begin{eqnarray}
\omega&=&\frac{1}{2\pi}\sum\limits_{k}\kappa_{k}\int\limits_{\Sigma}
\varepsilon^{lj}\partial_{l}\left(\varepsilon^{ij}\frac{(x^{i}-a^{i}_{k})}{|\vec{x}-\vec{a}_{k}|^{2}}\right)d^{2}x\nonumber\\
&&=
\frac{1}{2\pi}\sum\limits_{k}\kappa_{k}\int\limits_{\Sigma}\partial_{i}\frac{(x^{i}-a^{i})}{|\vec{x}-\vec{a}_{k}|^{2}}d^{2}x 
\end{eqnarray}
Moreover, we know that
\begin{eqnarray}
\partial_{i}\frac{(x^{i}-a^{i}_{k})}{|\vec{x}-\vec{a}_{k}|^{2}}= \partial_{i}\partial_{i}\ln{|\vec{x}-\vec{a}_{k}|}=2\pi\delta^{2}(\vec{x}-\vec{a}_{k}),
\end{eqnarray}
which leads to
\begin{eqnarray}
\omega=\sum\limits_{k} \kappa_{k},
\end{eqnarray}
indicating that $\omega$ counts the number of vortices and antivortices enclosed by $\gamma$ or, equivalently, those contained within $\Sigma$. This aligns with the essence of our invariant, where instead of vortices and antivortices, we consider points and antipoints.

%%%%%%%%%%%%%%%%%%%%%%%%%%%%%%%%%%%%%%%%%%%%%%%%%%%%%%%%%%%%%%%%%%%%%%%%
\section{Conclusions}
In this work, we have determined the explicit form of the generator for duality transformations that preserve the action of the Kalb-Ramond-Klein-Gordon model in $2+1$ dimensions, identifying it as a topological invariant. We applied the canonical quantization procedure, obtaining a formal representation of the algebra and duality generator in terms of the model's canonical operators. Our results reveal that the canonical algebra naturally emerges in a framework based on $2$-surfaces and signed points. Consequently, since the duality generator is a topological invariant, it admits a geometric interpretation as a link invariant that detects the number of signed points on a surface $\Sigma$. Alternatively, by interpreting signed points as sources of Faraday lines, the invariant can be understood as a measure of flux across the boundary of $\Sigma$, drawing a direct analogy to Gauss’s law in electrodynamics. Additionally, we established that this invariant provides a dual representation of the winding number.

This study presents the first explicit geometric representation of the invariant associated with massive $p$-form theories. While previous works such as Ref. \cite{ContrerasPRD} analyzed general duality generators for massive theories, they did not explicitly address the geometric invariant structure. Our findings align with those results but extend them by demonstrating that, in the case of massive $p$-forms, the invariant corresponds to the intersection number between $p$-surfaces and $q$-surfaces. Specifically, in our case, it arises from the intersection of a $2$-surface and a $0$-surface.

\section{acknowledgements}
E. C. is funded by the Beatriz Galindo contract BG23/00163 (Spain). E. C. acknowledge Generalitat Valenciana through PROMETEO PROJECT CIPROM/2022/13.

\bibliographystyle{unsrt}
\bibliography{references.bib}
%\begin{thebibliography}{99}
% SOLO LO DEJE PARA QUE APAREZCA ALGO, (TOCA REEMPLAZAR LA BIBLIOGRAFIA)

%\end{thebibliography}

\end{document}